\documentclass[lettersize,journal]{IEEEtran} 
\IEEEoverridecommandlockouts
\usepackage{amsthm,amsmath,amssymb,amsfonts,mathtools}
\usepackage{subcaption}
\usepackage{xcolor}
\usepackage{color,soul}

\DeclareMathOperator{\diag}{diag}

\usepackage{xspace} 
\usepackage{glossaries}
\setacronymstyle{long-short}
\newacronym[longplural = {reconfigurable intelligent surfaces}, shortplural={RISs}]{ris}{RIS}{reconfigurable intelligent surface} 
\newcommand{\ris}{\gls{ris}\xspace}
\newcommand{\riss}{\glspl{ris}\xspace}

\newacronym{ap}{AP}{access point} 
\newcommand{\ap}{\gls{ap}\xspace}

\newacronym[longplural = {user equipments}, shortplural={UEs}]{ue}{UE}{user equipment} 
\newcommand{\ue}{\gls{ue}\xspace}
\newcommand{\ues}{\glspl{ue}\xspace}

\newacronym[longplural = {neural networks}, shortplural={NNs}]{nn}{NN}{neural network} 
\newcommand{\nn}{\gls{nn}\xspace}
\newcommand{\nns}{\glspl{nn}\xspace}

\newacronym{csi}{CSI}{channel state information} 
\newcommand{\csi}{\gls{csi}\xspace}

\newacronym{mumiso}{MU-MISO}{multi-user multiple-input-single-output}

\newacronym{awgn}{AWGN}{additive white gaussian noise} 
\newcommand{\awgn}{\gls{awgn}\xspace}

\newacronym{ao}{AO}{alternating optimization}
\newcommand{\ao}{\gls{ao}\xspace}

\newacronym{bcd}{BCD}{block coordinate descent}
\newcommand{\bcd}{\gls{bcd}\xspace}

\newacronym{wmmse}{WMMSE}{weighted minimum mean square error}
\newcommand{\wmmse}{\gls{wmmse}\xspace}

\newacronym{pi}{PI}{power iteration}
\newcommand{\powi}{\gls{pi}\xspace}
\newcommand{\wmmsepi}{\wmmse-\powi}

\newacronym{dqnn}{DQNN}{deep quantization neural network}
\newcommand{\dqnn}{\gls{dqnn}\xspace}

\newacronym{dl}{DL}{deep learning}
\newcommand{\dl}{\gls{dl}\xspace}

\newacronym{aqewmmse}{AQE-WMMSE}{Auto-Quantization-Encoder and WMMSE beamforming updater}
\newcommand{\aqewmmse}{\gls{aqewmmse}\xspace}

\newacronym{aqe}{AQE}{Auto-Quantization-Encoder}
\newcommand{\aqe}{\gls{aqe}\xspace}

\newacronym{fc}{FC}{fully connected}
\newcommand{\fc}{\gls{fc}\xspace}

\newacronym{relu}{ReLU}{rectified linear unit}
\newcommand{\relu}{\gls{relu}\xspace}

\begin{document}

\title{Model-based Deep Learning for Joint RIS Phase Shift Compression and WMMSE Beamforming}

\author{\IEEEauthorblockN{Alexander James Fernandes and Ioannis Psaromiligkos} \\
\thanks{This work was supported in part by the Natural Science and Engineering Research Council of Canada under the Discovery Grant Program, the Vadasz Scholar McGill Engineering Doctoral Award, Calcul Québec (calculquebec.ca), and the Digital Research Alliance of Canada (alliancecan.ca). {Department of Electrical and Computer Engineering}, 
{McGill University}, Montreal, QC H3A 0E9, Canada. Email: 
alexander.fernandes@mail.mcgill.ca; ioannis.psaromiligkos@mcgill.ca. Accepted for publication in \emph{IEEE Wireless Communications Letters}. \copyright IEEE. DOI: 10.1109/LWC.2026.3683016}
}

\markboth{}%
{}

\maketitle

\begin{abstract}
A model-based \dl architecture is proposed for \ris-assisted multi-user communications to reduce the number of bits required for transmitting phase shift information from the \ap to the \ris controller.
The \ap computes the phase shifts and compresses them into a binary control message that is sent to the \ris controller for element configuration.
To help reduce beamformer mismatches caused by phase shift compression errors, the beamformer is updated with the actual (decompressed) \ris phase shifts.
By unrolling the iterative \wmmse algorithm within the wireless communication-informed \dl architecture, joint phase shift compression and \wmmse beamforming can be trained end-to-end.
Simulation results demonstrate that incorporating compression-aware beamforming significantly improves sum-rate performance, even when the number of control bits is lower than the number of \ris elements.
\end{abstract}

\glsresetall


\glsresetall

\section{Introduction}
The \ris has attracted significant interest as a means to control the wireless propagation environment~\cite{Wu2021}.
By jointly optimizing the beamformer and the phase shifts of the \ris passive reflective elements, wireless coverage is improved by enabling cooperative signal focusing and interference cancellation along the wireless links between the \ap and \ues.

Given its importance, joint optimization of the beamformers and \ris phase shifts is an active research topic \cite{Wu2019, Guo2020, Jin2024, Chen2024b, Xu2022, Li2024}.
Among the works using traditional optimization, \cite{Wu2019} introduced an \ao framework to minimize transmit power which was extended to physical‑layer security in \cite{Chen2024b}. \cite{Guo2020} proposed a \bcd approach to maximize the weighted sum‑rate, while \cite{Jin2024} developed a \wmmsepi algorithm with lower complexity than the earlier techniques.
Recent studies adopt \dl based strategies: \cite{Xu2022} employed a \dqnn for discrete \ris phase‑shift optimization, while \cite{Li2024} introduced ACFNet, a \dl architecture that compresses and decompresses \ris phase‑shift information between the \ap and \ris controller.

RISs are designed to be cost-effective and lack computing and sensing capabilities. 
As a result, computationally intensive tasks such as \csi estimation 
\cite{
Chen2021,
Fernandes2023,Fernandes2024,Fernandes2024a}, and joint optimization of the beamforming and phase shifts are performed at the \ap. 
The \ris controller therefore relies entirely on the \ap for phase shift reconfiguration.
However, transmitting the phase shift value for each \ris element introduces a communication overhead that scales linearly with the number of elements, creating a significant bottleneck in systems with large \riss~\cite{Zhang2021a}.
This necessitates compressing the phase shifts, which inevitably leads to decompression errors.
These errors represented as mismatches between the phase shifts assumed by the beamformer and those actually applied by the \ris, ultimately degrade the signal quality.
Therefore, joint beamforming and phase shift optimization must account for the information transfer constraints between the \ap and \ris controller.

Except for ACFNet~\cite{Li2024}, previous works do not consider the phase shift control message transmission constraints.
Although ACFNet compresses the \ris phase shifts, it does not account for the effective channel with the decompressed phase shifts when jointly optimizing beamforming and phase shifts.
Therefore, its compression gains come at the cost of decreasing the achievable sum-rate.
The conventional iterative algorithms in~\cite{Wu2019, Guo2020, Jin2024} were designed based on the wireless communication system model, while previous generic \dl methods in \cite{Xu2022,Li2024} try to automatically discover the model information from training data, but result in generalization errors.
These generalization errors indicate that there is room for improvement when designing a \dl architecture that considers phase shift control message transmission constraints.

In this paper, we propose the \aqewmmse \dl architecture  to jointly optimize the \ap beamformer and \ris phase shifts under a constraint on the number of control bits sent from the \ap to the \ris controller.
Specifically, our contributions are as follows:
\begin{itemize}
    \item Instead of individually transmitting phase shifts for each \ris element from the \ap to the \ris controller as done in~\cite{Wu2019, Guo2020, Jin2024, Chen2024b, Xu2022}, the \aqewmmse exploits the correlations between the phase shifts to reduce the number of bits required by extracting and combining relevant features into a single binary control message.
    \item In contrast to conventional iterative algorithms \cite{Wu2019,Guo2020,Jin2024}, we jointly compress the phase shifts and update the beamformer to maximize the achievable sum-rate using end-to-end \dl training.
    This is done by designing the \aqewmmse to use a trainable unrolled version of the iterative \wmmse algorithm~\cite{Monga2021a}.
    Compared to the generic \nns in~\cite{Xu2022, Li2024}, the hypothesis space of the \aqewmmse is smaller due to a model-based structure that reduces generalization error.
    \item The \aqewmmse accounts for the mismatch between the phase shifts assumed by the beamformer and those actually applied by the \ris by explicitly integrating the compression process into the joint beamforming and phase shift optimization. 
\end{itemize}

\section{System Model and Problem Statement}
\label{sec:systmodel-problem}
\subsection{System Model}

\begin{figure}[t]
    \centering
    \includegraphics[width=0.9\columnwidth]{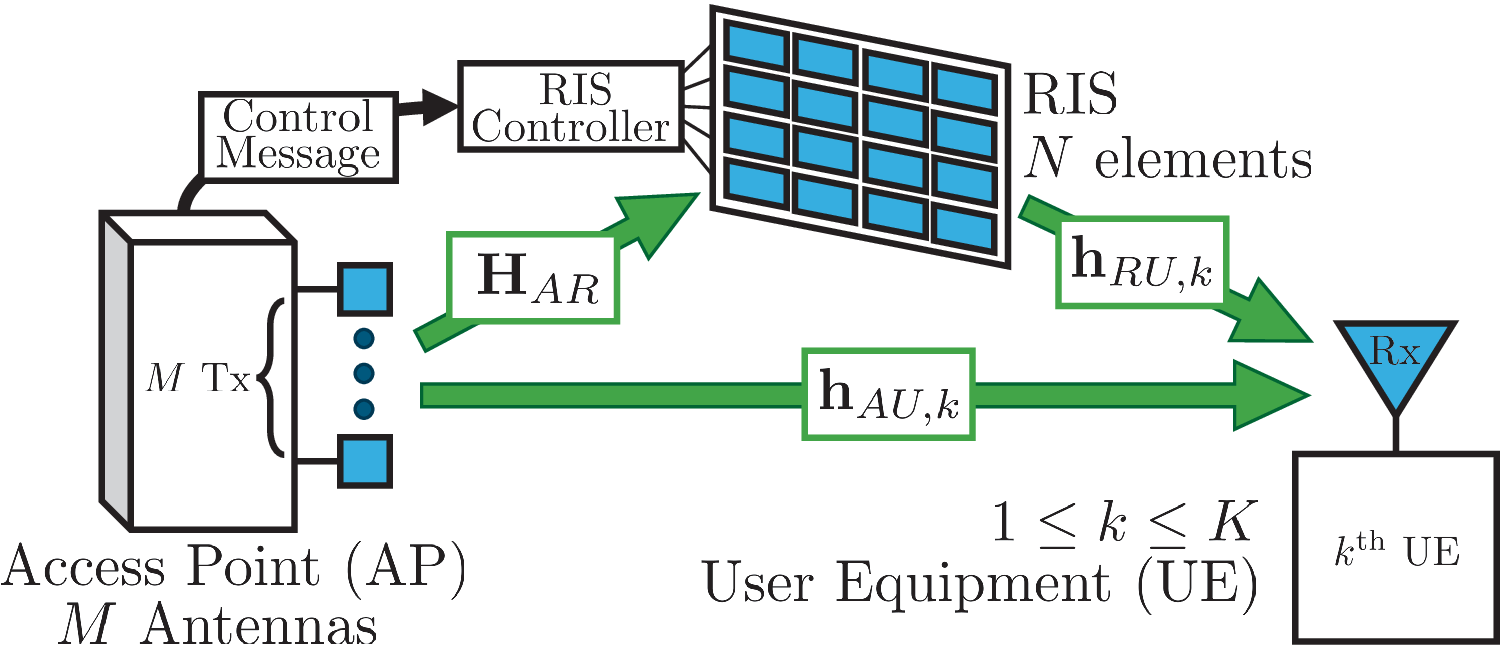}
    \caption{\ris-assisted communication system.}
    \label{fig:system-model}
\end{figure}

We consider the downlink of a narrowband \ris-assisted communication system, shown in Fig.~\ref{fig:system-model},  comprising an \ap with $M$ transmit antennas serving $K$ single-antenna \ues, and an \ris with $N$ elements.
The channels from the \ap to the \ris, from the \ris to the $k$-th UE, and from the \ap to the $k$-th \ue are denoted by $\textbf{H}_{AR} \in \mathbb{C}^{N \times M}$, $\textbf{h}_{RU,k} \in \mathbb{C}^{1 \times N}$, and $\textbf{h}_{AU,k} \in \mathbb{C}^{1 \times M}$, respectively.
All channels are assumed to be quasi-static and estimated\footnote{Techniques for channel estimation are explored in 
\cite{Chen2021}, 
\cite{Fernandes2023}, \cite{Fernandes2024}, and \cite{Fernandes2024a}.} at the \ap.
We also define the aggregate channel matrices $\textbf{H}_{RU} = [\textbf{h}_{RU,1}^T, \ldots, \textbf{h}_{RU,K}^T]^T \in \mathbb{C}^{K \times N}$, and $\textbf{H}_{AU} = [\textbf{h}_{AU,1}^T, \ldots, \textbf{h}_{AU,K}^T]^T \in \mathbb{C}^{K \times M}$.
The \ap beamforming vector for the $k$-th \ue is denoted by $\textbf{w}_k \in \mathbb{C}^{M \times 1}$.  
Collectively, $\textbf{w}_k$ satisfy a transmit power constraint $\sum_{k=1}^K \|\textbf{w}_k\|^2 \le P$, or $\|\textbf{W}\|^2_F \le P$, where $\textbf{W} = [\textbf{w}_1, \ldots, \textbf{w}_K] \in \mathbb{C}^{M \times K}$ is the beamforming matrix, and $\|\cdot\|_F$ denotes the Frobenius norm.
The \ris reflection coefficients are represented by the vector $\boldsymbol{\phi} = [e^{j\theta_1}, \ldots,  e^{j\theta_N}] \in \mathbb{C}^{1 \times N}$, where $\theta_{n} \in [-\pi, \pi)$ is the phase shift applied by the $n$-th element, $1 \le n \le N$.
The phase shifts are collected in the vector $\boldsymbol{\theta} = [\theta_1, \ldots, \theta_N]$.

The received signal $y_k \in \mathbb{C}$ at the $k$-th \ue, due to the transmission of symbols $s_l \in \mathbb{C}$, $1 \le l \le K$, to the \ues is:
\begin{align}
    \label{eq:sys-model}
    y_k &= \sum_{l=1}^K (\textbf{h}_{AU,k} + {\boldsymbol{\phi}\textbf{H}_k}) \textbf{w}_l s_l + n_k 
\end{align}
where $\textbf{H}_k = \diag({\textbf{h}_{RU,k}})\textbf{H}_{AR} \in \mathbb{C}^{N \times M}$ is the cascaded channel from the \ap to the $k$-th UE via the \ris, and $n_k \sim \mathcal{CN}(0, \sigma^2_k)$ is the \awgn at the $k$-th \ue.
The achievable rate to the $k$-th \ue is:

\begin{align}
    \label{eq:achievable-rate}
    R_k = \text{log}_2 \left( 1 + \frac{|(\textbf{h}_{AU,k} + \boldsymbol{\phi}\textbf{H}_k) \textbf{w}_k|^2}{\sum_{l=1,l \ne k}^K |(\textbf{h}_{AU,k} + \boldsymbol{\phi}\textbf{H}_k) \textbf{w}_l|^2 + \sigma_k^2} \right)
\end{align}

\subsection{Problem Statement}
Let $\mathcal{I}$ denote the available information at the \ap used to maximize the sum-rate of the system, which includes the \csi along with the optimal \ap beamformer and \ris phase shifts. 
We note that the \ris controller does not have access to $\mathcal{I}$.
To configure the \ris, the \ap computes and sends a binary control message, constrained to $B$ bits, to the \ris controller through a dedicated and error-free control link. 
The controller then decodes this message to determine the phase shifts to be used by the \ris. 
Formally, the process is defined by two functions: the \ap generates the binary control message $\boldsymbol{\psi} \in \mathcal{F} = \{\textit{0}, \textit{1}\}^B$ using $\boldsymbol{\psi} = f_{\text{c}}(\mathcal{I})$, while the \ris controller applies $\boldsymbol{\theta} = f_{\text{d}}(\boldsymbol{\psi})$ to decode $\boldsymbol{\psi}$ and obtain the phase shifts $\boldsymbol{\theta}$. 
Simultaneously, the \ap updates the beamforming matrix $\mathbf{W}$ to account for the phase shift compression using a third function $\mathbf{W}=f_{\text{w}}(\mathcal{I})$.

Our goal is to identify $f_{\text{c}}(\cdot)$, $f_{\text{w}}(\cdot)$, and $f_{\text{d}}(\cdot)$ that jointly maximize the achievable sum-rate over the $K$ \ues.
This problem\footnote{We note that, in the ideal case when the control link is not bandwidth-limited (i.e., $B \to \infty$),
problem (P1) reduces to the traditional joint beamforming and \ris phase shift optimization problem. 
In this scenario, $\textbf{W}_{\text{opt}} = f_{\text{w}}({\mathcal I})$ and $\boldsymbol{\theta}_\text{opt} = f_{\text{d}}(f_{\text{c}}({\mathcal I}))$, where $(\textbf{W}_{\text{opt}}, \boldsymbol{\theta}_{\text{opt}})$ denotes the optimal solution that can be obtained by \ao \cite{Wu2019}, \bcd \cite{Guo2020}, or \wmmsepi \cite{Jin2024}.} can be formulated as:
\begin{flalign}
    (\text{P1}): \quad \max_{f_{\text{c}}(\cdot), f_{\text{d}}(\cdot), f_{\text{w}}(\cdot)} \quad & \sum_{k=1}^K p_kR_k \notag \\
    \quad \textrm{s.t.} \quad \quad & \boldsymbol{\psi} = f_{\text{c}}({\mathcal I}) \in \mathcal{F} = \{\textit{0}, \textit{1}\}^B \notag \\
    \quad \quad & \boldsymbol{\theta} = f_{\text{d}}(\boldsymbol{\psi}) 
    \notag \\
    \quad \quad & \textbf{W} = f_{\text{w}}({\mathcal I}) \text{ with } \|\textbf{W}\|^2_F \le P \label{eq:P1}
\end{flalign}
where $p_k$ is the priority of serving the $k$-th UE.

\section{Proposed \dl Architecture}
\label{sec:prop-method}
In \cite{Xu2022,Li2024}, generic \nns are used to solve (P1).
In particular, the functions for beamforming $f_{\text{w}}(\cdot)$ and control message generation $f_{\text{c}}(\cdot)$ are combined into a single \nn $(\textbf{W}, \boldsymbol{\psi}) = f_{\text{w,c}}(\mathcal{I})$. 
In \cite{Li2024}, a second \nn $\boldsymbol{\theta} = f_{\text{d}}(\boldsymbol{\psi})$ then decodes the control message $\boldsymbol{\psi}$.
However, training the \nns $f_{\text{w,c}}(\cdot)$ and $f_{\text{d}}(\cdot)$ can be challenging.
This is because $f_{\text{w,c}}(\cdot)$ is prone to high generalization error due to its large hypothesis space~\cite[Section 5.2]{Goodfellow2016a}.
In addition, the beamformer matrix does not depend directly on the decoded phase shifts in $\boldsymbol{\theta}$. 
Instead,  $\textbf{W}$ and $\boldsymbol{\theta}$ are indirectly coupled via the aggregate loss $\mathcal{L}(\textbf{W}, \boldsymbol{\theta})$ and backpropagation of the corresponding gradients (see \cite[eq. (5) and (13)]{Xu2022} and \cite[eq. (7) and (10)]{Li2024}).

\begin{figure}[t]
    \centering
    \includegraphics[width=1\columnwidth]{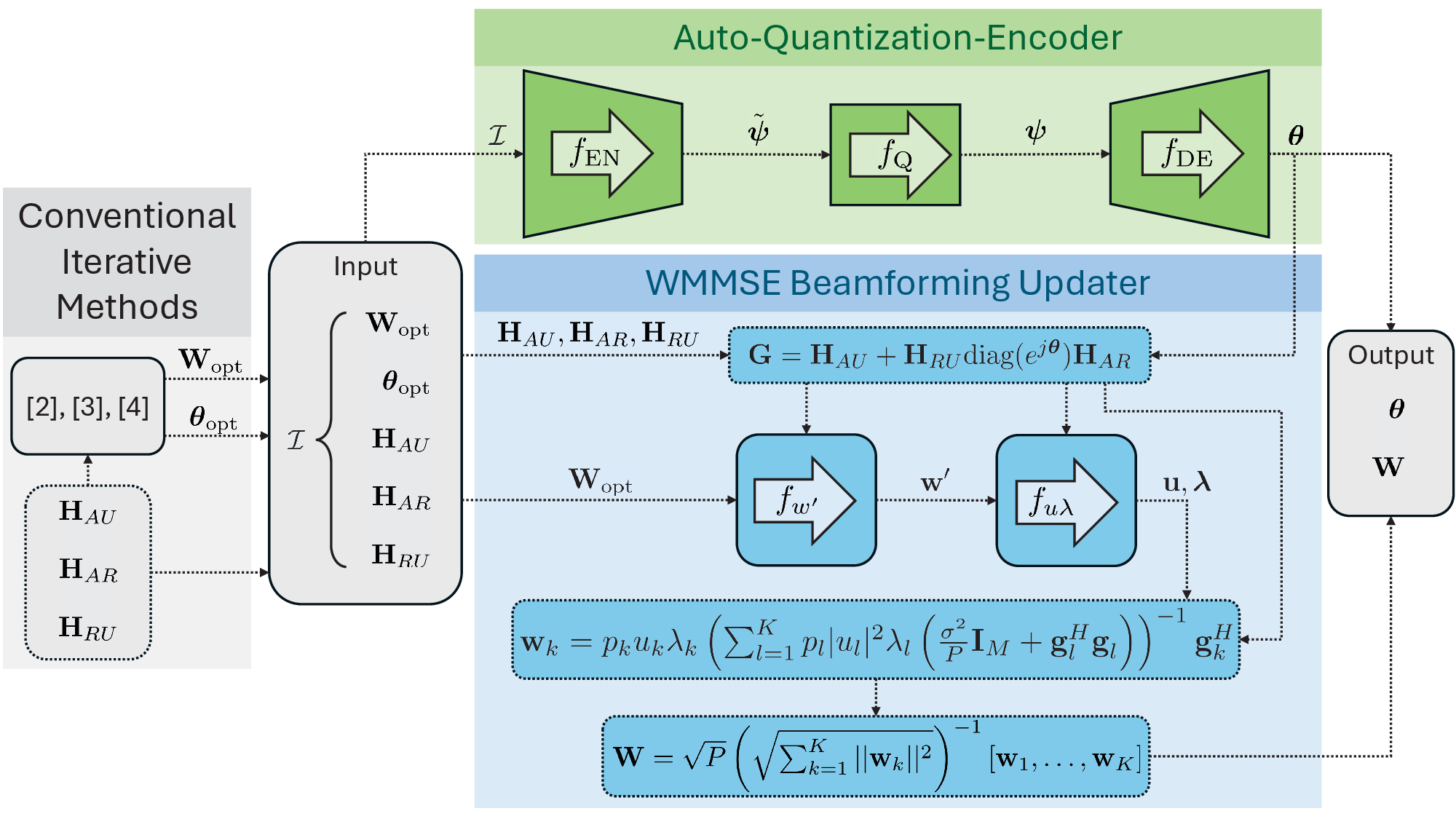}
    \caption{\dl architecture of the AQE-WMMSE network.}
    \label{fig:AQEWupdate-architecture}
\end{figure}

In this paper, we design the beamforming solution to explicitly take into account the decompressed phase shifts.
To avoid high generalization error, we propose a wireless communication informed \dl architecture \aqewmmse\footnote{Source code: https://github.com/AlexanderFernandes96/AQE-WMMSE} to solve (P1). 
As shown in Fig.~\ref{fig:AQEWupdate-architecture} it comprises two modules.
The first module, \aqe, is responsible for encoding the available information into a binary control message, and then decoding the control message into the \ris phase shifts.
The second module, \wmmse beamforming updater, is responsible for updating the beamforming matrix with the actual \ris reflection coefficients by using the phase shifts at the output of the \aqe.\footnote{We note that an identical copy of the \ris controller's \aqe decoder is used at the \ap to update $\textbf{W}$.}
To update the beamforming matrix we propose a model-based approach by unrolling the \wmmse algorithm as part of our \dl architecture.
The input of the \aqewmmse is all the information available at the \ap, $\mathcal{I} = \{\textbf{W}_{\text{opt}}, \boldsymbol{\theta}_{\text{opt}}, \textbf{H}_{AU}, \textbf{H}_{AR}, \textbf{H}_{RU}\}$, where ($\textbf{W}_{\text{opt}}, \boldsymbol{\theta}_{\text{opt}})$ is the solution to (P1) obtained with the \wmmsepi algorithm~\cite{Jin2024}.
The \wmmse beamforming updater module is directly coupled with the \aqe module since $\boldsymbol{\theta} = f_{\text{d}}(f_{\text{c}}(\mathcal{I}))$ and $\textbf{W} = f_{\text{w}}(\mathcal{I};\boldsymbol{\theta})$.\footnote{The notation $\textbf{W} = f_{\text{w}}(\cdot;\boldsymbol{\theta})$ signifies that the beamformer $\textbf{W}$ depends on the actual (decompressed) \ris phase shifts $\boldsymbol{\theta}$ at the \aqe output.}
The modules are also indirectly coupled as they are trained end-to-end using an aggregate loss $\mathcal{L}(\textbf{W}, \boldsymbol{\theta})$.

\subsection{Auto-Quantization-Encoder}
Similar to an autoencoder \nn, the \aqe network in Fig.~\ref{fig:AQEWupdate-architecture} consists of an encoder \nn $f_{\text{EN}}(\cdot)$ and a decoder \nn $f_{\text{DE}}(\cdot)$. 
In addition, the \aqe uses an intermediate quantization layer $f_{\text{Q}}(\cdot)$ between the encoder and decoder \nns \cite{Shlezinger2019,Shohat2019}.
In relation to (P1), the \nns perform: $\boldsymbol{\psi} = f_{\text{c}}(\mathcal{I}) = f_{\text{Q}}(f_{\text{EN}}(\mathcal{I}))$ and $\boldsymbol{\theta} = f_{\text{d}}(\boldsymbol{\psi}) = f_{\text{DE}}(\boldsymbol{\psi})$.

\subsubsection{Encoder}
The \nn $f_{\text{EN}}(\cdot)$ extracts $N_c$ phase features from the available information, $\Tilde{\boldsymbol{\psi}} = [\Tilde{\psi}_1, \ldots, \Tilde{\psi}_{Nc}] = f_{\text{EN}}(\mathcal{I})$.

\subsubsection{Quantizer}
The quantizer $\boldsymbol{\psi} = f_{\text{Q}}(\Tilde{\boldsymbol{\psi}}) \in \mathcal{F} = \{\textit{0}, \textit{1}\}^B$
associated with the binary constraint in (P1), transforms the phase features $\Tilde{\boldsymbol{\psi}}$ into the control message $\boldsymbol{\psi}$.
The function $f_{\text{Q}}(\cdot)$ employs a trainable scalar quantizer $Q(\cdot)$ with $D = 2^d$ levels~\cite{Shlezinger2019}. 
This quantizer is applied to each of the $N_c$ phase features, producing a $d$-bit quantized phase feature $\psi_n = Q(\tilde{\psi}_n)$ for $1 \leq n \leq N_c$. 
The $N_c$ quantized features are then concatenated into a $B = N_c d$-bit control message $\boldsymbol{\psi} = [\psi_1, \ldots, \psi_{N_c}]
\in \mathcal{F}$.

To train $Q(\cdot)$, we adopt the structure from \cite{Shohat2019}:
\begin{flalign}
    \label{eq:quantizer_softhard}
    Q(x) = \sum_{i=1}^{D-1} a_i q(c_i(x - b_i))
\end{flalign}
During training, to enable gradient backpropagation we relax the binary constraint in (P1) by using the continuous function $q(\cdot) = \tanh(\cdot)$.
On the other hand, during evaluation we use the discontinuous step function $q(\cdot)= \text{sign}(\cdot)$ to enforce the binary constraint.
The trainable parameters $a_i$, $b_i$, and $c_i$ represent the amplitude, shift, and slope of the $\tanh$ function.

\subsubsection{Decoder}
The decoder $f_{\text{DE}}(\cdot)$
reconstructs the $N$ \ris phase shifts in $\boldsymbol{\theta}$ from the control message $\boldsymbol{\psi}$.

\subsection{WMMSE Beamforming Updater}

To solve (P1), we update $\textbf{W}_{\text{opt}}$ using the \aqe output $\boldsymbol{\theta}$, i.e., $\textbf{W} = f_{\text{w}}(\mathcal{I};\boldsymbol{\theta})$.
To that end, we use the \wmmse algorithm~\cite{Shi2011} on the effective channel:
\begin{flalign}
    \textbf{G} &= \textbf{H}_{AU} + \textbf{H}_{RU} \diag(e^{j\boldsymbol{\theta}}) \textbf{H}_{AR} \label{eq:beamformingupdate-G}
\end{flalign}
where $\textbf{G}^T = [\textbf{g}_1, \ldots, \textbf{g}_K]^T \in \mathbb{C}^{M \times K}$. 
The \wmmse solution for the \ap beamforming vector of the $k$-th \ue, adapted from \cite[eq. (15)]{Shi2011} to the case of a single-antenna \ue, then replacing the Lagrange multiplier $\mu_k$ by the reciprocal SNR term $\sigma^2/P$ (similar to \cite[eq. (7) and (10)]{Jin2024}), is:
\begin{flalign}
    \textbf{w}_k &= p_k u_k \lambda_k \left(\sum_{l=1}^{K} p_l |u_l|^2 \lambda_l \left(\frac{\sigma^2}{P}\textbf{I}_{M} + \textbf{g}_l^H \textbf{g}_l\right)\right)^{-1} \textbf{g}_k^H \label{eq:beamformingupdate-w}
\end{flalign}
where $\textbf{u} = [u_1, \ldots, u_K] \in \mathbb{C}^{K}$, and $\boldsymbol{\lambda} = [\lambda_1, \ldots, \lambda_K]\in\mathbb{R}_+^K, \lambda_k \ge 0$ are the \wmmse receiver\footnote{The \wmmse receivers are used to estimate the data symbol at the $k$-th \ue as $\hat{s}_k=u_k^*y_k$ where $(\cdot)^*$ denotes complex conjugate.} and weight parameters, obtained through alternating optimizations, using $\textbf{W}_{\text{opt}}$ as the initial value.

To train the \aqewmmse end-to-end, we unroll the iterative algorithm in \cite{Shi2011} into the \wmmse beamforming updater depicted in {Fig.}~\ref{fig:AQEWupdate-architecture}.
We first propose to use the \nn $\textbf{w}' = f_{w'}(\textbf{W}_{\text{opt}}, \textbf{G})$ as an initialization stage, where $\textbf{w}' \in \mathbb{R}^{2MK}$ represents updating the optimal beamforming matrix $\textbf{W}_{\text{opt}}$ based on the actual channel $\textbf{G}$.
Then we propose the \nn $\textbf{u}, \boldsymbol{\lambda} = f_{u\lambda}(\textbf{w}', \textbf{G})$ to learn the \wmmse receiver and weight parameters \cite[eq. (5) and (13)]{Shi2011} to update the beamformer via \wmmse with (\ref{eq:beamformingupdate-w}).
Altogether, $f_{w'}$ initializes the beamformer, while $f_{u\lambda}$ and (\ref{eq:beamformingupdate-w}) implement the \wmmse algorithm. Compared to unrolling multiple iterations each requiring a matrix inversion in (\ref{eq:beamformingupdate-w}), we use a single iteration with $\textbf{u}$ and $\boldsymbol{\lambda}$ obtained by the \nn $f_{u\lambda}$.
Finally, the beamforming matrix is $\textbf{W} = \frac{\sqrt{P}} {\sqrt{\sum_{k=1}^K \|\textbf{w}_k\|^2}} [\textbf{w}_1, \ldots, \textbf{w}_K] \in \mathbb{C}^{M \times K}$ which satisfies the transmit power constraint.

\section{Simulation Results}
\label{sec:sim-results}
We simulate a uniform linear array antenna \ap with $M=4$, $K=3$ single-antenna UEs, and a uniform rectangular array \ris with $N = N_w \times N_h = 10 \times 10=100$. 
Transmission occurs over a bandwidth of 180 kHz, with a noise power spectral density of $-170$ dBm/Hz (for all \ues).
All channels are geometrically modelled with $m=10$ multi-paths \cite{Chen2021,Fernandes2024}.
The path loss is described by $\rho(d) = \rho_0(\frac{d}{d_0})^{-a}$, where $d$ is the distance between links, $a$ denotes the path loss exponent, and  $d_0 = 1$m, $\rho_0 = -30$dB.
With parameters:\footnote{$\rho_{AR} = 5.2481\times10^{-4}$, $\rho_{RU} = 4.3076$, and $\rho_{AU} = 3.3846\times10^{-5}$.} $d_{AR} = 50$m, $a_{AR} = 2.8$, $d_{RU} = 2$m, $a_{RU} = 2.8$, $d_{AU} = 50.04$m, $a_{AU} = 3.5$.
To ensure the \ap-\ris-\ue link is stronger than the \ap-\ue link, the channels are generated then scaled w.r.t. the path losses as: $\textbf{H}_x = \sqrt{\rho_x} \tilde{\textbf{H}} / ||\tilde{\textbf{H}}_x||_2$, $x \in\{AU, AR, RU\}$.

\subsection{Model Parameters}
\label{sec:architecture-parameters}
The blocks in the \aqewmmse are structured as follows.\footnote{
All complex variables are represented by real vectors by separating the real and imaginary components.
}

$f_{\text{EN}}$: For a fair comparison, the model complexity in multiplications is kept on the same order as the DNN in~\cite{Xu2022}\footnote{Although the encoder does not need to be large, the DNN in \cite{Xu2022} can help to mitigate the impact of imperfect \csi.}, with slightly more parameters to handle the inputs $\textbf{W}_{\text{opt}}$ and $\boldsymbol{\theta}_{\text{opt}}$.
By denoting $H=N+2KM$, we use five \fc \nn layers of $32H$, $16H$, $8H$, $4H$, and $N_c$ neurons. 
The first four layers have the following components in order: \fc \nn, \relu, dropout (probability of $0.5$), then batch normalization. 
The last layer consists only of the \fc \nn.
The complexity is $O(N^2(64K + 64M) + N(192KM + 4Nc + 128K^2M + 128KM^2) + 256K^2M^2 + 1344KM + 8KMN_c) \approx O(N^2(64K + 64M))$.

$f_{\text{Q}}$: The scalar quantizer is set to 1-bit, i.e., $D=2$.
Consequently, each encoder output $\psi_n$ corresponds to one bit of the control message, 
for a total of $B=N_c$ control bits.
The parameters are initialized to $a_1 = \frac{\pi}{2}$, $b_1 = 0$ and $c_1 = 1$.
The slope $c_1$ is non-trainable and is increased by $c_1^{(i)} = 1.005c_1^{(i-1)}$ after each epoch $i$~\cite{Shohat2019}.

$f_{\text{DE}}$: 
The decoder\footnote{The decoder at the \ris controller is designed with less complexity than the encoder: $O(2N^2 + NN_c)$, which is lower than the ACFNet decoder \cite{Li2024} $O(6N(2N_h-1)(N_w-1) + 2N(12 + N_c)) \approx O(12N^2 + 2NN_c)$.
It is also easier to update the parameters in the decoder than in the encoder due to the vanishing gradients from the flat regions of the quantizer \cite{Goodfellow2016a,Shohat2019}.} is an \fc \nn with three linear layers each having $N$ neurons.
The first two layers use \relu, while the output layer has no activation function.

$f_{w'}$ and $f_{u\lambda}$: both are four-layer \fc \nns.
Each of the first three layers has $4KM$ neurons with a \relu activation function followed by a dropout layer ($0.5$ probability).
The last layer of $f_{w'}$ and $f_{u\lambda}$ comprises $2KM$ neurons for $\textbf{w}'$ and $3K$ neurons for $(\textbf{u},\boldsymbol{\lambda})$, respectively.
No activation function is applied to the outputs of $f_{w'}$.
The first $2K$ outputs of $f_{u\lambda}$ for $\textbf{u}$ use no activation function, while the absolute value function is applied to the remaining $K$ outputs of $f_{u\lambda}$ to ensure non-negative elements for $\boldsymbol{\lambda}$.
The complexity of $f_{w'}$ is $O(M^2(56K^2 + 8NK) + M(8K^2N + 4KN))$ and $f_{u\lambda}$ is $O(M^2(48K^2 + 8NK) + M(8NK^2 + 4NK + 12K^2))$ (lower than the $O(M^3)$ matrix inverse in (\ref{eq:beamformingupdate-w})).

\subsection{Training Procedure}
\label{sec:architecture-training}

We generate $64{,}000$, $16{,}000$, and $20{,}000$ samples for training, validation, and testing, respectively.
In all experiments, we use a batch size of $L=512$ with the maximum number of epochs set to $1{,}000$ saving the model with the lowest validation loss.
The learning rate is initialized to $0.001$ and reduced by a factor of $0.8$ when the validation loss does not decrease for 20 consecutive epochs, until reaching a minimum value of $0.00005$.
The loss function being minimized for each batch is:
$\mathcal{L}(\textbf{W}, \boldsymbol{\theta}) = -\sum_{l=1}^L \sum_{k=1}^K p_k R_k$.
When comparing loss curves, the loss is normalized by dividing by the number of training or validation samples.
For fair comparison, we use a priority $p_k=1$ for all $K$ \ues.
The achievable sum-rates presented are averages of five independent trials.

\subsection{Benchmarks}

\textit{Upper Bound}: We use the \wmmsepi algorithm \cite{Jin2024} with 100 iterations, to obtain the optimal \ap beamforming matrix $\textbf{W}_{\text{opt}}$ and continuous valued \ris phase shifts $\boldsymbol{\theta}_{\text{opt}}$.

\textit{Benchmark 1} (\dqnn \cite{Xu2022}): A \dl network that quantizes the phase shifts to $\pm a_1$, with each quantized value encoded using one bit.
Therefore, the resulting control message has length $B = N$.
The same \dl network is also used to obtain the beamforming matrix based on the \csi \cite[eq. (10), Fig. 2]{Xu2022}.

\textit{Benchmark 2} (ACFNet \cite{Li2024}): A \dl network that compresses the bits of the control message.
The DNN layer from \cite[Fig. 2]{Xu2022} is used as the input of the ACFNet in~\cite[Fig. 2]{Li2024} with $32H$, $16H$, $8H$, $4H$, and $2M+N$ neurons ($H=N+2MK$), to obtain the beamforming matrix and phase shift vector before compression.
For a fair comparison, compression is set to $B=N_c$ bits by removing the Policy Network in \cite[eq. (8) and Fig. 2]{Li2024} and the phase shifts at the \ris are continuous valued by removing $Q_2$ in \cite[eq. (9) and Fig. 2]{Li2024}.

\textit{Benchmark 3} (linQ \cite{Shlezinger2019,Shohat2019}):
Given $\textbf{W}_{\text{opt}}$ and $\boldsymbol{\theta}_{\text{opt}}$, we implement linear analog combining and quantization \cite{Shlezinger2019,Shohat2019} 
(\nns are single layer matrix multiplication with bias and no activation function) 
to solve (P1) with $\boldsymbol{\theta} = f_\text{D}(f_{\text{Q}}(f_\text{C}(\boldsymbol{\theta}_{\text{opt}})))$ and $\textbf{W} = f_W(\textbf{W}_{\text{opt}})$.
$f_\text{C}$ and $f_\text{D}$ compress/decompress the phase shifts, and $f_W$ updates the beamformer with complexity $O(NN_c)$, $O(NN_c)$, $O(M^2)$, respectively.

\textit{Benchmark 4} (\aqe): The proposed \aqe \nn is trained without the \wmmse Beamforming Updater module.
The output of this network is: $\textbf{W} = \textbf{W}_{\text{opt}}$ and $\boldsymbol{\theta} = f_{\text{DE}}(f_{\text{Q}}(f_{\text{EN}}(\mathcal{I})))$.

All benchmarks use the quantization scheme $f_{\text{Q}}(\cdot)$ outlined in Sec.~\ref{sec:architecture-parameters} to construct the control message.
In the \dqnn, ACFNet, linQ, and \aqe benchmark methods, the beamforming matrix $\textbf{W}$ does not depend on the actual phase shifts $\boldsymbol{\theta}$ used by the \ris controller.
The \dqnn and ACFNet benchmarks share the same complexity bottleneck: the DNN for \csi feature extraction adopted from \cite{Xu2022} with complexity $O(N^2(64K + 64M) + N(2768KM + 128K^2M + 128KM^2) + 2832K^2M^2)  \approx O(N^2(64K + 64M)) \approx O(f_{\text{EN}})$.

\subsection{Results}
\begin{figure*}[t]
    \centering
    \begin{subfigure}[b]{0.245\textwidth}
        \includegraphics[width=\textwidth]{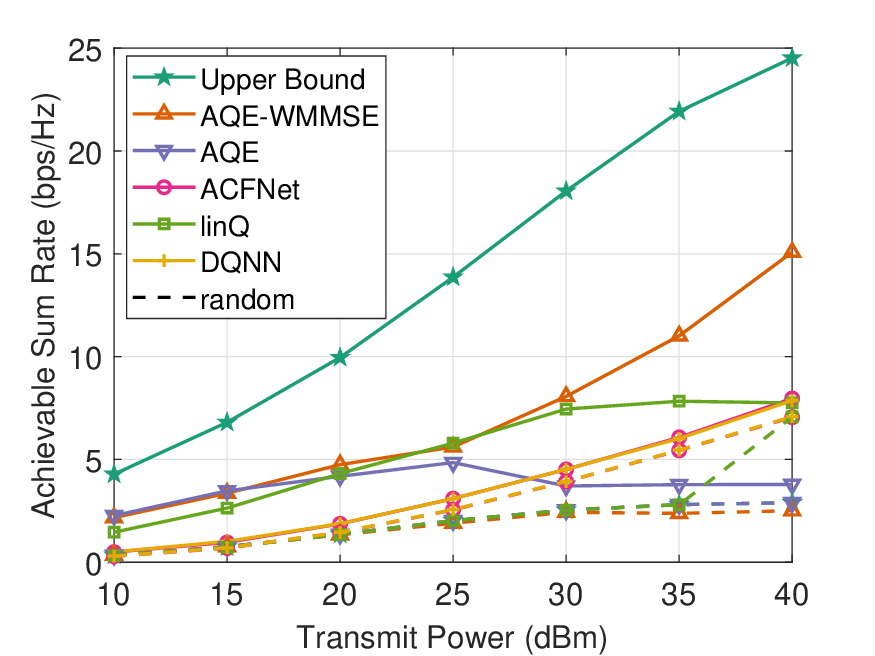}
        \caption{}
        \label{fig:RateVsTxPower_100bits}
    \end{subfigure}
    \begin{subfigure}[b]{0.245\textwidth}
        \includegraphics[width=\textwidth]{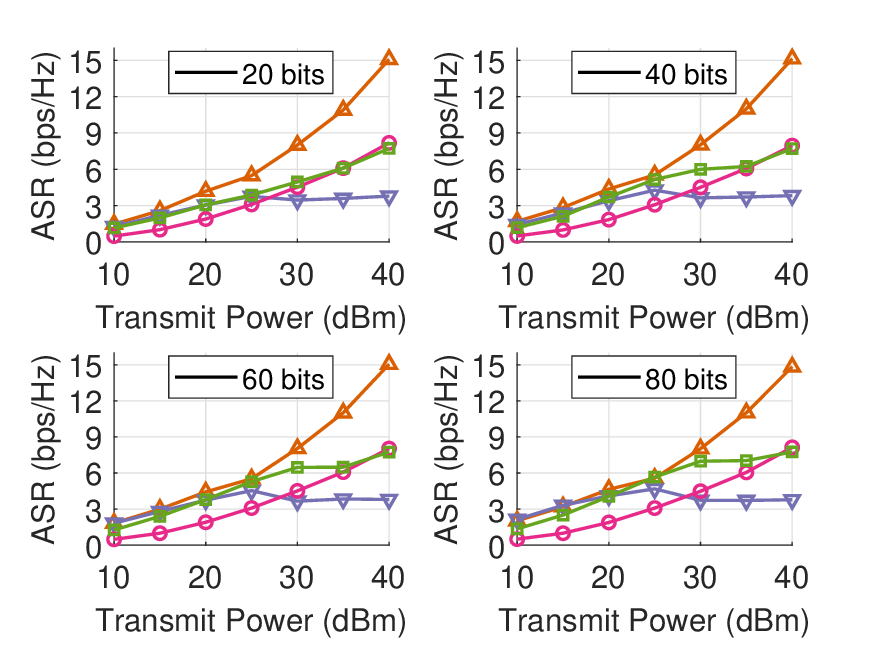}
        \caption{}
        \label{fig:RateVsTxPower_90vs10bits}
    \end{subfigure}
    \begin{subfigure}[b]{0.245\textwidth}
        \includegraphics[width=\textwidth]{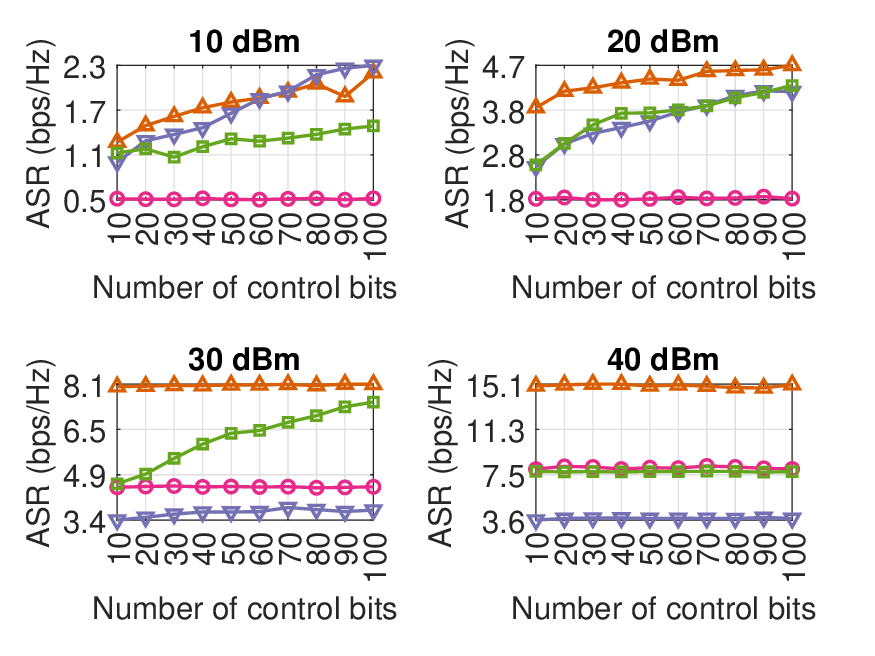}
        \caption{}
        \label{fig:RateVsBits}
    \end{subfigure}
    \begin{subfigure}[b]{0.245\textwidth}
        \includegraphics[width=\textwidth]{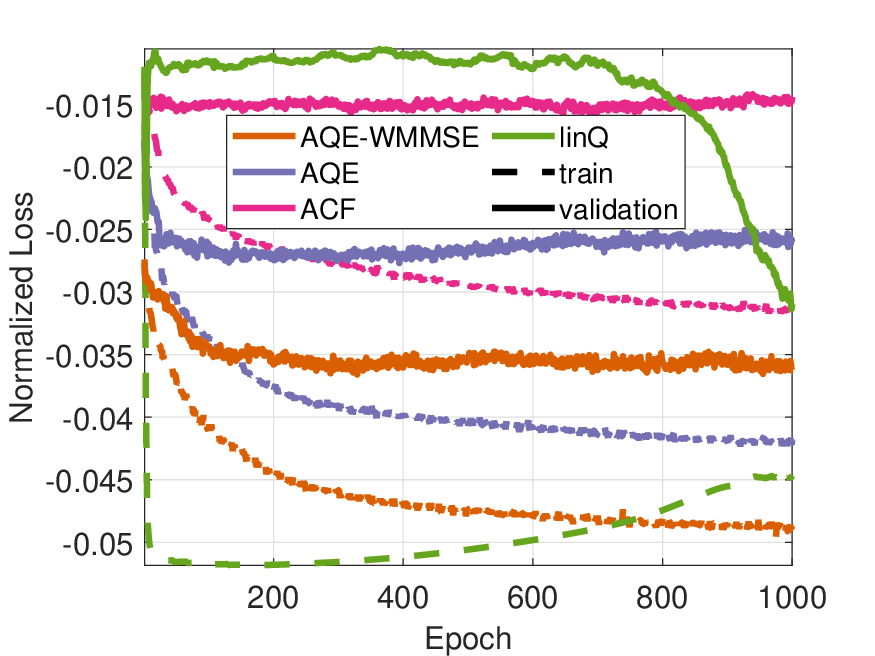}
        \caption{}
        \label{fig:loss}
    \end{subfigure}
    \caption{Achievable sum rate (ASR) vs transmit power $P$ with $N_c = N = B = 100$ bits in (a), and $B=20, 40, 60$, and $80$ in (b). ASR vs $B$ in (c). Normalized loss curves vs training epochs for $P = 20$ dBm, $B = 40$ in (d).}
    \label{fig:Rate}
\end{figure*}

In Table \ref{tab:Parameters} we show the number of parameters of the considered models for various control message sizes $B$.
\begin{table}[htbp]
    \centering
    \begin{tabular}{c|c|c|c}
         Model & $B=10$ & $B=40$ & $B=100$ \\ \hline
         AQE-WMMSE & 16,539,458 & 16,557,368 & 16,593,188 \\
         AQE & 16,523,729 & 16,541,639 & 16,577,459 \\
         ACFNet & 16,175,065 & 16,190,005 & 16,219,885\\
         linQ & 2,713 & 8,743 & 20,803 \\
         DQNN & --- & --- & 16,067,055
    \end{tabular}
    \caption{Num. of parameters ($N=100$, $M=4$, $K=3$).}
    \label{tab:Parameters}
\end{table}

In {Figs.}~\ref{fig:RateVsTxPower_100bits} and~\ref{fig:RateVsTxPower_90vs10bits} we compare the achievable sum-rate for the \aqewmmse against the benchmarks as a function of the transmit power $P$ for number of control bits $B = N_c = N$.
We also present the performance of each method in {Figs.}~\ref{fig:RateVsTxPower_100bits} when the
control message fails to reach the RIS controller, in which case
the RIS uses random phase shifts.
In Fig.~\ref{fig:RateVsBits} we show the performance as a function of $B$ for different values of $P$.
Fig.~\ref{fig:loss} shows normalized loss curves vs training epochs, which illustrates that the \aqewmmse has a lower validation loss than the benchmarks, even as the soft-quantizer's slope $c_1$ approaches a hard-quantizer (at epoch 1000: $c_1^{(1000)}=146.576$).
At low transmit power, the \aqewmmse, \aqe, and linQ obtain higher achievable sum rates than the ACFNet and DQNN, which perform similar to their corresponding random phase shifts.
At higher transmit power, the \aqe, which uses $\textbf{W}_{\text{opt}}$ as the beamformer, fails to optimally compress the phase shifts.
In contrast, the \aqewmmse performs significantly better demonstrating the necessity of joint beamforming and phase shift compression.

\begin{figure}[t]
    \centering
    \begin{subfigure}[b]{0.24\textwidth}
        \includegraphics[width=\textwidth]{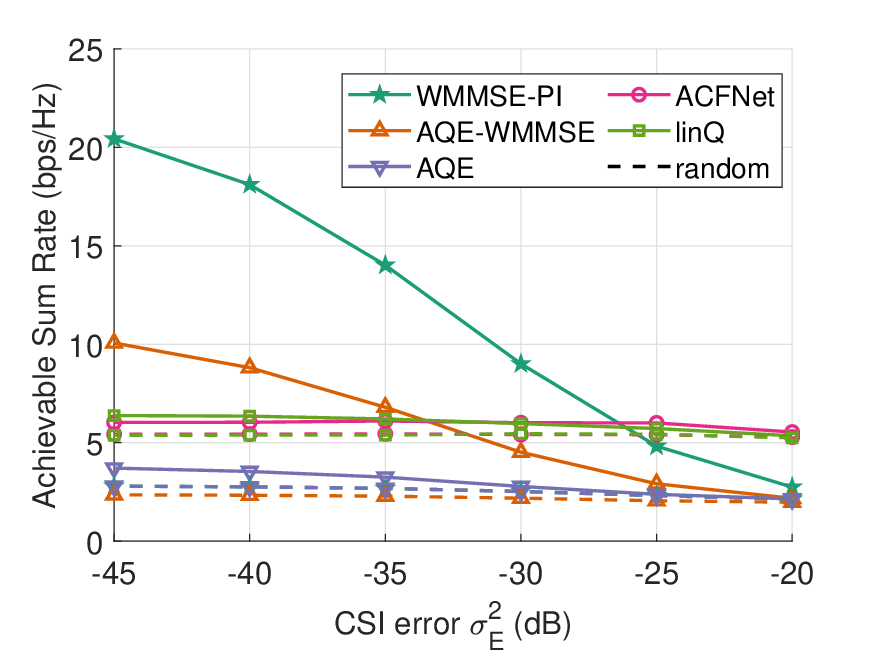}
        \caption{Imperfect \csi}
        \label{fig:ImperfectCSI}
    \end{subfigure}
    \begin{subfigure}[b]{0.24\textwidth}
        \includegraphics[width=\textwidth]{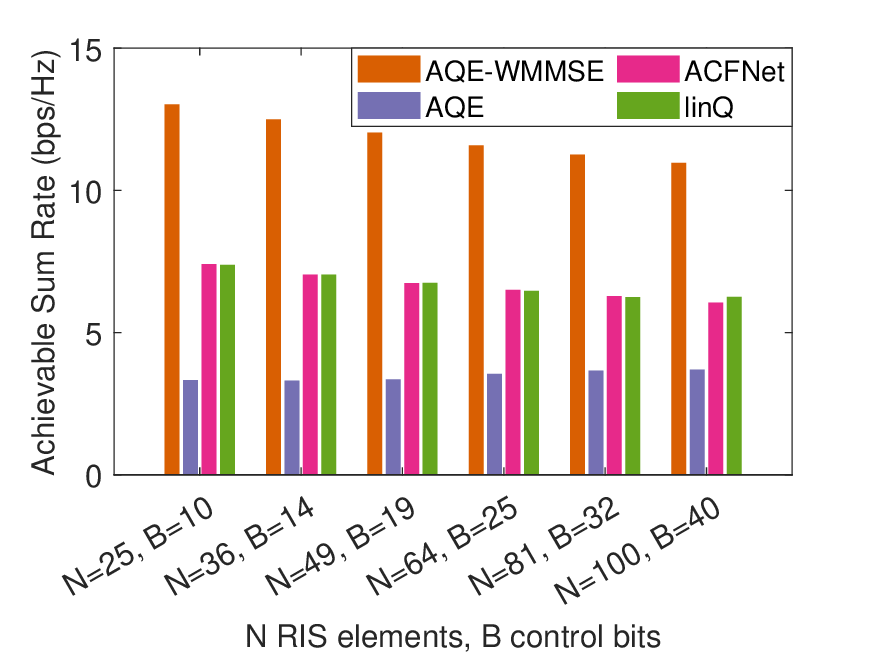}
        \caption{Square \ris ($N_h = N_w$)}
        \label{fig:varyN}
    \end{subfigure}
    \caption{Sum rate as a function of (a) \csi error ($N=100$, $B=40$), and (b) $N$ and $B$ (perfect \csi). $P=35$dBm.}
    \label{fig:4}
\end{figure}

Fig.~\ref{fig:ImperfectCSI} shows the impact of imperfect \csi.
The \csi errors are modelled as an additive error matrix $\textbf{E}_x$ with i.i.d. elements $\mathcal{CN}(0, \sigma^2_E||\textbf{H}_x||_F^2)$, $x \in\{AU, AR, RU\}$.
As the \csi error increases, the performance of \wmmsepi (with imperfect \csi) and \aqewmmse decreases; in contrast, the benchmarks perform close to their random phase shifts for all \csi errors.
Fig.~\ref{fig:varyN} shows the performance for different combinations of $N$ and $B$, with channels normalized as explained in Sec. \ref{sec:sim-results}.

\section{Conclusion}
\label{sec:conclusion}
A wireless communication informed \dl architecture was proposed to jointly compress the \ris phase shifts and update the \ap beamforming matrix by reconstructing the phase shift information into a control message. 
Numerical results showed that updating the beamforming matrix based on the effective channel obtains a higher achievable sum-rate than recent methods in the literature when the number of bits in the control message is lower than the number of \ris elements.

\bibliographystyle{IEEEtran}
\bibliography{RIS_PhaseShiftFeedbackReferences}

@article{Wu2021,
archivePrefix = {arXiv},
arxivId = {2007.02759},
author = {Wu, Qingqing and Zhang, Shuowen and Zheng, Beixiong and You, Changsheng and Zhang, Rui},
doi = {10.1109/TCOMM.2021.3051897},
eprint = {2007.02759},
issn = {15580857},
journal = {IEEE Trans. Commun.},
month = {May.},
number = {5},
pages = {3313--3351},
publisher = {Institute of Electrical and Electronics Engineers Inc.},
title = {{Intelligent Reflecting Surface-Aided Wireless Communications: A Tutorial}},
volume = {69},
year = {2021}
}

@article{Wu2019,
archivePrefix = {arXiv},
arxivId = {1810.03961},
author = {Wu, Qingqing and Zhang, Rui},
doi = {10.1109/TWC.2019.2936025},
eprint = {1810.03961},
issn = {1536-1276},
journal = {IEEE Trans. Wirel. Commun.},
month = {Nov.},
number = {11},
pages = {5394--5409},
publisher = {Institute of Electrical and Electronics Engineers Inc.},
title = {{Intelligent Reflecting Surface Enhanced Wireless Network via Joint Active and Passive Beamforming}},
volume = {18},
year = {2019}
}

@article{Guo2020,
archivePrefix = {arXiv},
arxivId = {1912.11999},
author = {Guo, Huayan and Liang, Ying-Chang and Chen, Jie and Larsson, Erik G.},
doi = {10.1109/TWC.2020.2970061},
eprint = {1912.11999},
issn = {1536-1276},
journal = {IEEE Trans. Wirel. Commun.},
month = {May.},
number = {5},
pages = {3064--3076},
publisher = {Institute of Electrical and Electronics Engineers Inc.},
title = {{Weighted Sum-Rate Maximization for RIS Aided Wireless Networks}},
volume = {19},
year = {2020}
}

@article{Jin2024,
author = {Jin, Weijie and Zhang, Jing and Wen, Chao-Kai and Jin, Shi and Li, Xiao and Han, Shuangfeng},
doi = {10.1109/TWC.2023.3336742},
issn = {1536-1276},
journal = {IEEE Trans. Wirel. Commun.},
month = {Jul.},
number = {7},
pages = {6968--6982},
publisher = {Institute of Electrical and Electronics Engineers Inc.},
title = {{Low-Complexity Joint Beamforming for RIS-Assisted MU-MISO Systems Based on Model-Driven Deep Learning}},
volume = {23},
year = {2024}
}

@article{Xu2022,
archivePrefix = {arXiv},
arxivId = {2111.06555},
author = {Xu, Wangyang and Gan, Lu and Huang, Chongwen},
doi = {10.1109/TCCN.2021.3128605},
eprint = {2111.06555},
issn = {2332-7731},
journal = {IEEE Trans. on Cogn. Commun. Netw.},
month = {Jun.},
number = {2},
pages = {694--706},
publisher = {Institute of Electrical and Electronics Engineers Inc.},
title = {{A Robust Deep Learning-Based Beamforming Design for RIS-Assisted Multiuser MISO Communications With Practical Constraints}},
volume = {8},
year = {2022}
}

@article{Li2024,
author = {Li, Zhicheng and Shen, Hong and Xu, Wei and Chen, Dong and Zhao, Chunming},
doi = {10.1109/LWC.2023.3342900},
issn = {21622345},
journal = {IEEE Wirel. Commun. Lett.},
number = {3},
pages = {766--770},
publisher = {IEEE},
title = {{Deep Learning-Based Adaptive Phase Shift Compression and Feedback in IRS-Assisted Communication Systems}},
volume = {13},
year = {2024}
}

@article{Chen2021,
author = {Chen, Xiao and Shi, Jianfeng and Yang, Zhaohui and Wu, Liang},
doi = {10.1109/LWC.2021.3054004},
issn = {2162-2337},
journal = {IEEE Wirel. Commun. Lett.},
month = {may},
number = {5},
pages = {996--1000},
publisher = {Institute of Electrical and Electronics Engineers Inc.},
title = {{Low-Complexity Channel Estimation for IRS-Enhanced Massive MIMO}},
volume = {10},
year = {2021}
}

@article{Fernandes2023,
archivePrefix = {arXiv},
arxivId = {2301.05263},
author = {Fernandes, Alexander James and Psaromiligkos, Ioannis},
doi = {10.1109/LWC.2023.3288517},
eprint = {2301.05263},
issn = {2162-2337},
journal = {IEEE Wirel. Commun. Lett.},
month = {Oct.},
number = {10},
pages = {1697--1701},
title = {{Channel Estimation for RIS-Assisted Full-Duplex MIMO With Hardware Impairments}},
volume = {12},
year = {2023}
}

@inproceedings{Fernandes2024,
address = {Washington DC, USA},
author = {Fernandes, Alexander James and Psaromiligkos, Ioannis},
booktitle = {IEEE Veh. Technol. Conf. (VTC)},
doi = {10.1109/VTC2024-Fall63153.2024.10757482},
month = {Oct.},
pages = {1--6},
publisher = {IEEE},
title = {{Joint Estimation of Direct and RIS-assisted Channels with Tensor Signal Modelling}},
year = {2024}
}

@article{Fernandes2024a,
author = {Fernandes, Alexander James and Psaromiligkos, Ioannis},
doi = {10.1109/OJCOMS.2024.3506481},
issn = {2644-125X},
journal = {IEEE Open J. Commun. Soc.},
number = {November},
pages = {7668--7684},
publisher = {IEEE},
title = {{Tensor Signal Modeling and Channel Estimation for RIS-Assisted Full-Duplex MIMO}},
volume = {5},
year = {2024}
}

@article{Shlezinger2019,
archivePrefix = {arXiv},
arxivId = {1807.08305},
author = {Shlezinger, Nir and Eldar, Yonina C. and Rodrigues, Miguel R. D.},
doi = {10.1109/TSP.2019.2935864},
eprint = {1807.08305},
issn = {1053-587X},
journal = {IEEE Trans. Signal Process.},
month = {Oct.},
number = {20},
pages = {5223--5238},
publisher = {Institute of Electrical and Electronics Engineers Inc.},
title = {{Hardware-Limited Task-Based Quantization}},
volume = {67},
year = {2019}
}

@inproceedings{Shohat2019,
author = {Shohat, Matan and Tsintsadze, Georgee and Shlezinger, Nir and Eldar, Yonina C.},
booktitle = {IEEE Int. Conf. on Acoust., Speech and Signal Process. (ICASSP)},
doi = {10.1109/ICASSP.2019.8682704},
isbn = {978-1-4799-8131-1},
issn = {15206149},
month = {May.},
pages = {3912--3916},
publisher = {IEEE},
title = {{Deep Quantization for MIMO Channel Estimation}},
volume = {2019-May},
year = {2019}
}

@article{Shi2011,
author = {Shi, Qingjiang and Razaviyayn, Meisam and Luo, Zhi-Quan and He, Chen},
doi = {10.1109/TSP.2011.2147784},
issn = {1053-587X},
journal = {IEEE Trans. Signal Process.},
month = {Sep.},
number = {9},
pages = {4331--4340},
title = {{An Iteratively Weighted MMSE Approach to Distributed Sum-Utility Maximization for a MIMO Interfering Broadcast Channel}},
volume = {59},
year = {2011}
}

@book{Goodfellow2016a,
author = {Goodfellow, Ian and Bengio, Yoshua and Courville, Aaron},
publisher = {MIT Press},
title = {{Deep Learning}},
url = {https://www.deeplearningbook.org/},
year = {2016}
}

@article{Zhang2021a,
archivePrefix = {arXiv},
arxivId = {2012.10736},
author = {Zhang, Hongliang and Di, Boya and Han, Zhu and Poor, H. Vincent and Song, Lingyang},
doi = {10.1109/LWC.2021.3058637},
eprint = {2012.10736},
issn = {2162-2337},
journal = {IEEE Wirel. Commun. Lett.},
month = {May.},
number = {5},
pages = {1098--1102},
publisher = {Institute of Electrical and Electronics Engineers Inc.},
title = {{RIS Assisted Multi-User Communications: How Many Reflective Elements Do We Need?}},
volume = {10},
year = {2021}
}

@article{Monga2021a,
archivePrefix = {arXiv},
arxivId = {1912.10557},
author = {Monga, Vishal and Li, Yuelong and Eldar, Yonina C.},
doi = {10.1109/MSP.2020.3016905},
eprint = {1912.10557},
issn = {1053-5888},
journal = {IEEE Signal Process. Mag.},
month = {Mar.},
number = {2},
pages = {18--44},
publisher = {Institute of Electrical and Electronics Engineers Inc.},
title = {{Algorithm Unrolling: Interpretable, Efficient Deep Learning for Signal and Image Processing}},
volume = {38},
year = {2021}
}

@article{Chen2024b,
author = {Chen, Zhen and Guo, Yeyong and Zhang, Peichang and Jiang, Hao and Xiao, Yuhang and Huang, Lei},
doi = {10.1109/LCOMM.2024.3427010},
issn = {1089-7798},
journal = {IEEE Commun. Lett.},
month = {nov},
number = {11},
pages = {2493--2497},
publisher = {Institute of Electrical and Electronics Engineers Inc.},
title = {{Physical Layer Security Improvement for Hybrid RIS-Assisted MIMO}},
volume = {28},
year = {2024}
}

\end{document}